\documentclass[graybox,natbib,nosecnum]{svmult}
\bibpunct{(}{)}{;}{a}{}{,} 

\pdfoutput=1   

\usepackage{mathptmx}       
\usepackage{helvet}         
\usepackage{courier}        
\usepackage{type1cm}        

\usepackage{makeidx}         
\usepackage{graphicx}        
\usepackage{multicol}        
\usepackage[bottom]{footmisc}
\usepackage[normalem]{ulem}	
\usepackage{hyperref}  

\usepackage{soul}   

\usepackage{amsmath}
\usepackage{amssymb}
\usepackage{amstext}
\usepackage{graphicx}
\usepackage{epstopdf}
\usepackage{epsfig}
\usepackage{url}

\makeindex             

\begin{document}

\title*{Atmospheric Retrieval of Exoplanets}
\author{Nikku Madhusudhan}
\institute{Institute of Astronomy, University of Cambridge, Madingley Road, Cambridge CB3 0HA, UK 
\email{nmadhu@ast.cam.ac.uk}} 
%
%
\maketitle

\abstract{
Exoplanetary atmospheric retrieval refers to the inference of atmospheric properties of an exoplanet given an observed spectrum. The atmospheric properties include the chemical compositions, temperature profiles, clouds/hazes, and energy circulation. These properties, in turn, can  provide key insights into the atmospheric physicochemical processes of exoplanets as well as their formation mechanisms. Major advancements in atmospheric retrieval have been made in the last decade, thanks to a combination of state-of-the-art spectroscopic observations and advanced atmospheric modeling and statistical inference methods. These developments have already resulted in key constraints on the atmospheric H$_2$O abundances, temperature profiles, and other properties for several exoplanets. Upcoming facilities such as the JWST will further advance this area. The present chapter is a pedagogical review of this exciting frontier of exoplanetary science. The principles of atmospheric retrievals of exoplanets are discussed in detail, including parametric models and statistical inference methods, along with a review of key results in the field. Some of the main challenges in retrievals with current observations are discussed along with new directions and the future landscape.}

\section{Introduction} 
\label{sec:intro}

A spectrum of an exoplanet provides a window into its atmosphere. A spectrum encodes information regarding the various interconnected physicochemical processes and properties of the atmosphere which are revealed through their influence on the radiation emerging through the atmosphere before reaching the observer. These properties include the chemical composition, temperature structure, atmospheric circulation, clouds/hazes, all of which leave their imprints on the spectrum. Given an observed spectrum, the challenge is to disentangle these various components. This is the goal of `atmospheric retrieval' -- to retrieve the atmospheric properties of an exoplanet from an observed spectrum. The retrieved properties can in turn provide insights into the various atmospheric physical and chemical processes as well as into their formation history. While the introduction of atmospheric retrieval methods to exoplanetary science is a relatively recent and independent development \citep{madhu2009,madhu2011a}, alternate techniques have been in wide usage in the context of Earth-based remote sensing \citep{rodgers2000} and retrievals of solar system planets \citep{irwin2008}. 

What differentiates exoplanetary atmospheric retrieval from solar system applications is the uniquely challenging nature of observing exoplanetary atmospheres. Firstly, unlike solar system planets, observed exoplanetary spectra are inherently disk-averaged over the spatially unresolved planet. Secondly, given their astronomical origins well beyond the solar system, exoplanetary spectra are naturally substantially fainter, and hence of much lower signal-to-noise (SNR), compared to solar-system objects. Thirdly, any complementary in situ measurements or a priori knowledge possible in the solar system are unavailable for exoplanetary atmospheres. Finally, the parameter space of exoplanetary atmospheres is substantially wider than that of solar system planets. For example, while the equilibrium temperatures of most solar system planets lie below 300 K those of exoplanets extend up to $\sim$3000 K. Similarly large ranges are natural in all other atmospheric parameters and processes - gravities, chemical compositions, circulation patterns, degree and type of insolation, etc, implying enormous complexity and diversity in exoplanetary atmospheres far beyond those experienced in the solar system. The combination of these various factors make exoplanetary atmospheres enormously more challenging to study compared to those of solar system objects, and necessitate substantially more robust techniques for atmospheric modeling and retrieval to make the best use of the limited spectral data available. 

The origins of atmospheric retrieval techniques for exoplanets were motivated by the `degeneracy problem' faced by early atmospheric observations. Initial molecular detections were claimed based on few channels of infrared photometry or low-resolution spectrophotometry with low SNR \citep[e.g.][albeit some of these datasets have since been revised substantially]{barman2007,tinetti2007,grillmair2008,swain2008a}, such that the spectral features were rarely discernible to the eye. Similarly, temperature inversions were claimed in hot Jupiters based on broadband photometric observations \citep[e.g][]{knutson2008,knutson2009,burrows2007,burrows2008}. These inferences were made using a limited set of forward models containing the putative molecules and assumed temperature profiles that qualitatively matched the data. While the number of free parameters in the forward models typically far exceeded the number of data points then available, the number of models compared against the data were rather limited. This approach left vast areas of parameter space unexplored and degeneracies between various model parameters unknown, thereby providing little statistical basis to the claimed detections. The desire to provide a statistically robust framework to derive atmospheric properties of exoplanets from such low resolution data gave birth to the idea of atmospheric retrieval for exoplanets \citep{madhu2009}. Atmospheric retrieval techniques have since advanced greatly in tandem with parallel advancements in atmospheric observations of exoplanets.

In the present chapter we present a pedagogical review of exoplanetary atmospheric retrieval. We first present an overview of the key principles of atmospheric retrieval. We then discuss two primary components of retrieval methods, namely, parametric forward models and statistical inference methods. We then discuss key results in the field from retrievals of state-of-the-art observations. We conclude with a discussion of key issues in this area and the future landscape. 

\begin{figure*}[t]
\centering
\includegraphics[width=\textwidth]{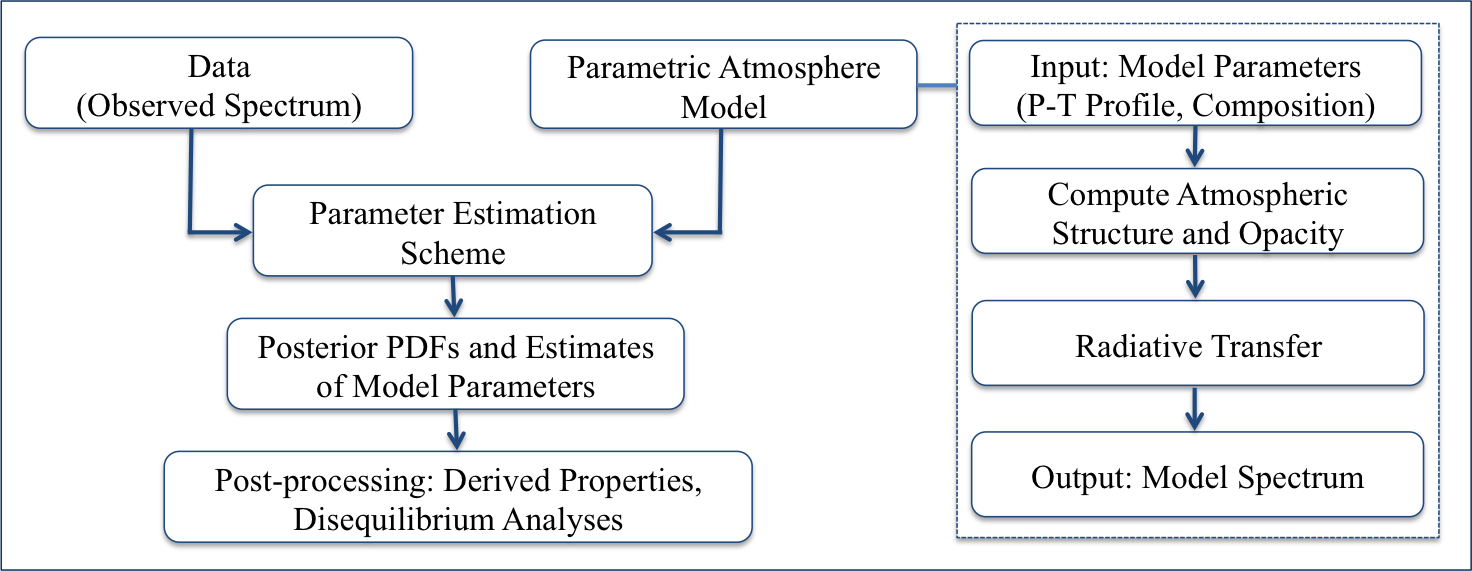}
\caption{Schematic of atmospheric retrieval. Given an observed spectrum and a parametric model of a planetary atmosphere, a parameter estimation method is used to derive the model parameters. The components of a typical atmospheric model are shown on the right. The free parameters typically correspond to the pressure-temperature (P-T) profile and the composition, including the chemical abundances and cloud/haze properties, depending on the datasets. The statistical inference and parameter estimation methods used in contemporary retrieval codes typically allow computation of full posterior probability density functions (PDFs) of the model parameters given a data set, a typical output  shown in Fig.~\ref{fig:retrieval}. These PDFs can also be used to compute PDFs of derived quantities such as elemental abundance ratios from those of molecular abundances. In recent advancements retrieval codes are also being coupled with self-consistent equilibrium models to place constraints on departures from radiative-convective and chemical equilibria \citep{gandhi2018}.}
\label{fig:schematic}
\end{figure*}

\begin{figure*}[t]
\centering
\includegraphics[width=\textwidth]{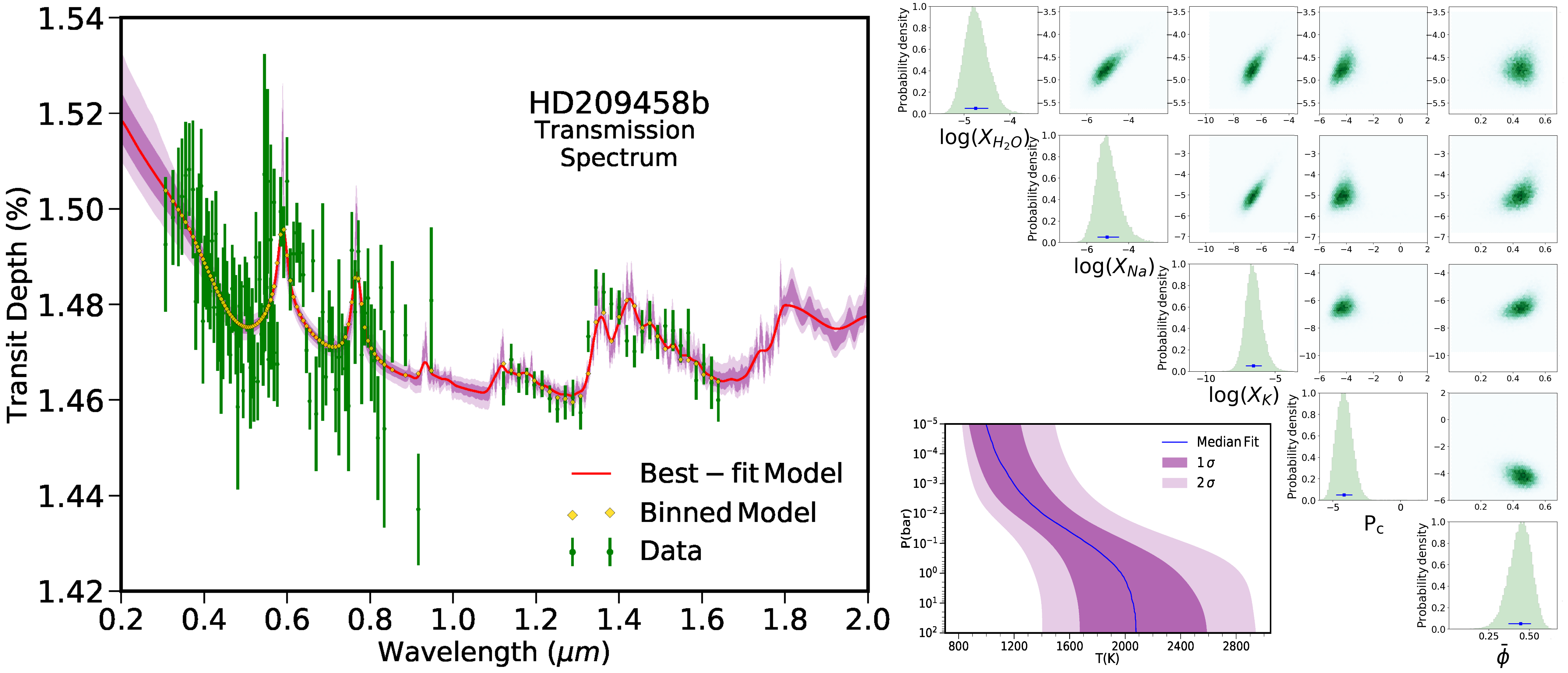}
\caption{ Example of atmospheric retrieval for a transmission spectrum of the hot Jupiter HD 209458b. The left panel shows an observed spectrum in green along with the model fit and significance contours in purple. The right panel shows the posterior probability distributions of the retrieved compositions and the retrieved pressure-temperature profile.}
\label{fig:retrieval}
\end{figure*}

\section{Overview of Atmospheric Retrieval}  
\label{sec:overview}

In its simplest form, `retrieval' is synonymous with fitting an atmospheric model to an observed spectrum and estimating the model parameters along with uncertainties. A schematic of atmospheric retrieval for exoplanets is shown in Fig.~\ref{fig:schematic} with an example output in Fig.~\ref{fig:retrieval}, and a list of extant retrieval codes in the literature is shown in  Table~\ref{table:retrieval_codes}. A parameter estimation problem requires three key components: 

\begin{enumerate}
\item{A reliable data set, in the present case an atmospheric spectrum.} 
\item{An accurate model, in the present case a parametric atmosphere model.}
\item{A suitable parameter estimation method.}
\end{enumerate}

All these components started becoming accessible for studying exoplanetary atmospheres only about a decade ago. The history of exoplanetary atmospheric retrieval is essentially the history of key developments in each of these aspects. In this section, we discuss the basic principles of atmospheric retrieval for exoplanetary atmospheres.

\subsection{Self-consistent models vs parametric retrieval models} 

At the outset it is important to distinguish between the two paradigms for forward modeling of  spectra of exoplanetary atmospheres - self-consistent models and parametric models used for atmospheric retrieval. Self-consistent models refer to models where the physicochemical properties and processes of the atmosphere are assumed to be known a priori. For example, a one-dimensional equilibrium model  to compute thermal emission from a planet typically assumes a plane parallel atmosphere under the constraints of hydrostatic equilibrium, chemical equilibrium, local thermodynamic equilibrium (LTE), and radiative-convective equilibrium. The inputs to such a model are typically the system properties, or `fixed parameters', such as the planetary bulk properties (e.g. mass, radius, and/or gravity), orbital properties (e.g. separation), and the spectrum of the host star irradiating the planet. Such a model also assumes an elemental composition of the planetary atmosphere, apart from other ancillary parameters (e.g. day-night energy redistribution efficiency, presence of clouds, etc.). Given these input parameters and the equilibrium assumptions such a model computes the radiative transfer in the atmosphere to generate an output spectrum for the desired viewing geometry. Self-consistent models span a wide range of complexity depending on the applications, ranging from 1D equilibrium models to 3D General Circulation Models (GCMs) as well as  models with non-equilibrium chemistry with varying levels of detail. A more detailed description of self-consistent models can be found in a number of recent sources\citep[e.g.][]{burrows2008,fortney2008,heng2017b,gandhi2017}. 

While self-consistent models are invaluable for investigating detailed atmospheric processes under controlled conditions they are limited in their capability to robustly interpret observed spectra. Firstly, self-consistent models, by definition,  assume that the physics and chemical compositions underlying the atmospheric processes are known. On the contrary, the novelty and diversity of exoplanetary atmospheres means that very little is known about them a priori and that the atmospheres could diverge substantially from the self-consistent model assumptions. Secondly, self-consistent models are not amenable to detailed parameter estimation methods due to their long computation times. For example, given the intricate degeneracies between the various atmospheric properties a typical model fitting to a spectral dataset could involve 10$^5$ - 10$^6$ model evaluations, which are prohibitive even for the simplest of self-consistent models, e.g. 1-D equilibrium models. The desired solution to interpret observed spectra of exoplanetary atmospheres is therefore a modeling paradigm which allows for (a) fast model evaluations, and (b) a parametrization of atmospheric properties that captures properties of self-consistent models as well as possible deviations thereof. For example, such a model should be able to replicate the temperature profiles, compositions, and spectra of self-consistent models for the same system parameters but also be flexible enough to model profiles and compositions that are not in equilibrium. 

Retrieval methods, on the other hand, employ parametric forward models to extract the atmospheric properties from observed spectra. An inspection of a nominal cloud-free self-consistent model reveals that the atmospheric spectrum is governed mainly by the pressure-temperature ($P$-$T$) profile and the chemical composition of the atmosphere, both of which are calculated self-consistently under equilibrium conditions in such models, with the additional possibility of clouds. As such self-consistent models have no free parameters but only fixed parameters. The key idea in retrieval models is to use parametric forms for the $P$-$T$ profile and the composition, which leads to two key advantages. Firstly, computing the $P$-$T$ profile and chemical compositions are the most time consuming steps in self-consistent models. Therefore, parameterizing both these properties substantially shortens the computing time of a model. Secondly, there is no longer the need to assume equilibrium conditions because the $P$-$T$ profile and chemical compositions are estimated directly from the data. These parametric models can be coupled with statistical parameter estimation methods to efficiently explore the model parameter space thereby allowing to formally fit the models to a given dataset and estimate the $P$-$T$ profiles, compositions, and other parameters, e.g. clouds/hazes, etc. This functionality is the backbone of atmospheric retrieval. 

An atmospheric retrieval code has two components: 1. a parametric model to compute the atmospheric spectrum for given atmospheric parameters, 2. an  optimization algorithm, i.e., a statistical inference method to sample the model parameter space given the data. For a given dataset the optimization technique explores the model parameter space in search of models fitting the data and in the process creates posterior probability distributions of all the model parameters. In the following sections we discuss both these components in detail, followed by a review of results from retrievals in the literature. 

\begin{table*}
\centering
\caption{Exoplanetary Atmospheric Retrieval Codes}
\label{table:retrieval_codes}
\begin{tabular}{p{4cm} l l l}
\\
\hline
\hline
\rule{0pt}{3ex}  Code Name/Author$^{a}$ & Forward Model & Inference Method$^{b}$ & References\\ [1ex]
\hline
\\ [-1.5ex]
\rule{0pt}{1ex}\\
Madhusudhan \& Seager & \parbox{2.8cm}{Primary Transit\\Secondary Eclipse} & Grid Search & \citet{madhu2009} \\[2ex] \\
Madhusudhan et al & \parbox{2.8cm}{Primary Transit\\Secondary Eclipse} & MCMC & \parbox{5cm}{\citet{madhu2010} \\ \citet{madhu2011a,madhu2014b}}\\[2ex] \\
CHIMERA & \parbox{2.8cm}{Primary Transit\\ Secondary Eclipse\\Direct Imaging} & \parbox{2.4cm}{OE, BMC, MCMC \\and Multinest NS} & \citet{line2013, line2014, todorov2016}\\[2ex] \\
NEMESIS & \parbox{2.8cm}{Primary Transit\\ Secondary Eclipse} & OE &  \parbox{5cm}{\citet{barstow2017} \\ \citet{lee2012}}\\[2ex] \\
Benneke \& Seager & Primary Transit & Multinest NS & \citet{benneke2013}\\[2ex] \\
$\mathcal{T}$-REx & \parbox{2.8cm}{Primary Transit\\ Secondary Eclipse} & Multinest NS, MCMC &  \parbox{5cm}{\citet{waldmann2015a,waldmann2015b}} \\[2ex] \\
HELIOS-R & \parbox{2.8cm}{Direct Imaging \\ Secondary Eclipse} & Multinest NS & \parbox{2.8cm}{\citet{lavie2017} \\ \citet{oreshenko2017}} \\[2ex] \\
ATMO & \parbox{2.8cm}{Primary Transit\\ Secondary Eclipse} & MCMC &  \parbox{5cm}{\citet{wakeford2017,evans2017}} \\[2ex] \\
BART & \parbox{2.8cm}{Primary Transit\\Secondary Eclipse} & MCMC & \parbox{5cm}{\citet{cubillos2015}\\ \citet{blecic2015}} \\[2ex]\\
POSEIDON & \parbox{2.8cm}{Primary Transit} & \parbox{2.4cm}{Multinest NS} & \citet{macdonald2017} \\[2ex] 
HyDRA & \parbox{2.8cm}{Secondary Eclipse} & \parbox{2.4cm}{Multinest NS} & \citet{gandhi2018} \\ [2ex] 
\\[-1.5ex]
\hline
\end{tabular} \\[1ex]

$^a$ Here we only list codes reported in published works that have been used on actual observations of exoplanetary spectra.  \\
$^b$ The statistical inference and parameter estimation method used in the retrieval code, e.g., Markov chain Monte Carlo (MCMC), Bootstrap Monte Carlo (BMC), Optimal Estimation (OE), and Nested Sampling (Multinest NS). See text for discussion on the different methods. \\

\end{table*}

\section{Models for Atmospheric Retrieval} 
\label{sec:models} 

The forward models and their parametrization for retrieval depend on the nature of the atmospheric observations in question. Exoplanetary atmospheric spectra used for retrievals have been observed in primarily three configurations: (a) transmission spectra of transiting exoplanets, (b) emission spectra of transiting  exoplanets, and (c) emission spectra of directly-imaged planets\footnote{Recently, retrieval codes are also being built for reflection spectra of directly imaged planets that can be obtained with future space-based facilities \citep{lupu2016}. However, we do not discuss these models here since they have so far only been used on simulated data and not on observed spectra of known exoplanets.}. Depending on the observing mode and geometry the observed spectrum is sensitive to a certain region of the atmosphere thereby requiring the corresponding model set-up and free parameters as discussed below. 

\subsection{General Framework and Free Parameters}
The goal of a parametric forward model used for retrievals is to compute a model atmospheric spectrum for the required observational configuration given the properties of the atmosphere as free parameters. So, there are two components to a forward model, as shown in Fig.~\ref{fig:schematic}: (1) computing the structure of the atmosphere, i.e. profiles of pressure ($P$), temperature ($T$), density ($\rho$), concentrations ($f_i$) of individual chemical species, cloud/haze profile, if any, etc., and (2) computing the radiative transfer for the given atmospheric structure. Here we briefly describe the general principles for computing the atmospheric structure. The radiative transfer for each observing configuration will be discussed in following subsections. 

Common to all the models used in retrieval are some general assumptions about the atmospheric structure. The atmospheres are assumed to be generally spherically symmetric, in hydrostatic equilibrium, and in local thermodynamic equilibrium (LTE). The opacities for the radiative transfer are usually computed in a line-by-line manner \citep{madhu2009,madhu2014b,line2013,benneke2013} but some codes use correlated-K opacities \citep{lee2012, lavie2017}. The $P$-$T$ profile and chemical compositions are free parameters in the models. Given a $P$-$T$ profile, the profiles of $P$, $T$, and $\rho$ as a function of radial distance ($r$) are determined using the assumption of hydrostatic equilibrium and the ideal gas equation of state. Given the parametric chemical composition, the mean molecular weight and total number density ($n$) are also determined. Once all these quantities are determined, the remaining task is to determine the radiation field emerging from the system for the given geometry by considering the appropriate scheme for radiative transfer in the atmosphere. 

The parameters for forward models used in retrievals correspond to three broad properties: chemical composition, $P$-$T$ profile, and clouds/hazes. The chemical composition of the atmosphere is represented by the volume mixing ratios ($f_i$) of the species, e.g. number density of each species relative to the total number density, implying as many free parameters as the number of species (typically between 4 and 10). Usually, for H$_2$-rich species the prominent absorbers such as H$_2$O, CO, CH$_4$, CO$_2$, Na, K, etc., are included. Additionally, the mixing ratios are typically assumed to be uniform in the region of the atmosphere probed by the observations. The $P$-$T$ profile is represented by one of two parametric $P$-$T$ profiles used in the retrieval literature, either the 6-parameter profile prescribed by \citet{madhu2009} or the 5-parameter profile reported by \citet{guillot2010} which, for example, was used in \citet{line2013}. Both profiles have been shown to reproduce characteristic $P$-$T$ profiles in planetary atmospheres, though the first one offers more flexibility at the cost of an extra parameter \citep{line2016b}. Additionally, the model can include opacity due to the presence of clouds or hazes in the atmosphere as well as the possibility of inhomogeneous clouds \citep{benneke2012,kreidberg2015,barstow2017,line2016a,macdonald2017}, adding $\gtrsim3$ more free parameters. In total, a typical parametric model has $\gtrsim$10 free parameters. 

\subsection{Transmission Spectra of Transiting Planets}

A transmission spectrum is observed when a planet transits in front of the host star. In this geometry some light from the host star passes through the atmosphere at the day-night terminator region of the planet before reaching the observer. This light is subjected to extinction, i.e. absorption and/or scattering, in the planetary atmosphere. This modified stellar spectrum when subtracted from the original stellar spectrum obtained out of transit gives the extinction spectrum of the planetary atmosphere. A `transmission spectrum' is represented as the extinction spectrum normalized by the original stellar spectrum and is essentially the transit depth as a function of the wavelength \citep{seager2000,seager2010}. The computation of a transmission spectrum at different levels of complexity, analytically and numerically, can be found in various works \citep{brown2001,hubbard2001,seager2000,desetangs2008a,miller-ricci2009,fortney2010,deWit2015,betremieux2015,macdonald2017,robinson2017}. Here we provide a brief outline. 

In its simplest form the transmission spectrum can be expressed as 
\begin{equation}
\Delta_\lambda = \Big(\frac{R_{p,\lambda}}{R_{s,\lambda}}\Big)^2 = \frac{2}{{R_s}^2}\int_0^{R_{max}}rdr~(1-e^{-\tau_\lambda(r)}), 
\end{equation}
where $\lambda$ is the wavelength, $R_p$ and $R_s$ are the planetary and stellar radii, respectively, and $R_{max}$ is the maximum height of the observable atmosphere typically set at a reasonably high value. $r$ is the impact parameter or height in the atmosphere perpendicular to the direction of the ray, and $\tau(r)$ is the slant optical depth through the chord traversed by a ray at the impact parameter $r$. The optical depth encountered by a ray as it traverses a chord is governed by the opacity from the planetary atmosphere at various pressures, temperatures, and chemical compositions on its path. The transmission spectrum is a cumulative effect of the opacity encountered by all the rays within the planetary atmosphere before reaching the observer. 

\subsection{Thermal Emission Spectra of Transiting Planets}

The transit geometry allows observations of thermal emission from the dayside atmosphere of the planet at opposition, also known as secondary eclipse or occultation. The occultation depth gives the planet star flux ratio as  
\begin{equation}
\frac{f_p}{f_s} = \frac{F_{out} - F_{in}}{F_{out}}, 
\end{equation}
where $F_{out}$ and $F_{in}$ are the fluxes from the system observed out of eclipse and during eclipse, respectively, and $f_p$ and $f_s$ are the planetary and stellar fluxes. The observed flux from a spherical body of radius $R$ at a distance $d$ is related to the specific intensity of radiation at its surface by $f_\lambda = \pi I_\lambda R^2/d^2$. Thermal emission models are used to compute the planetary spectrum given by $I_{\lambda,p}$ whereas the stellar  spectrum is obtained from standard libraries of stellar models. The observation of $f_p/f_s$ means that the distance to the system need not be known and, since the planetary radius is already know from transit, only $I_{\lambda,p}$ needs to be computed in models. 

Models of thermal emission spectra used in retrievals generally assume a 1-D plane parallel geometry \citep[e.g.][]{seager2010,madhu2009,line2013}. Consider a ray with spectral intensity $I_{0,\lambda}$ originating from a layer in the atmosphere at an optical depth $\tau$ with a direction cosine $\mu$. The specific intensity of the ray as it emerges out from the top of the atmosphere is given by 
\begin{equation}
I_\lambda(\tau,\mu) = I_{0,\lambda}(\tau,\mu)e^{-\tau_\lambda/\mu} - \frac{1}{\mu} \int_\tau^0 S_\lambda e^{-t/\mu} dt.  
\end{equation}
In this notation, $\tau = 0$ at the topmost layer of the atmosphere and increases inward such that $\tau \rightarrow \infty$ for the deepest layers. $S_\lambda$ is the source function, which for an atmosphere in Local Thermodynamic Equilibrium (LTE) is the Planck function. A ray generated in a deeper layer of the atmosphere traverses through the layers above before escaping the atmosphere and reaching the observer. On its way out photons in the ray are absorbed or enhanced depending on the opacity and temperature profile in the atmosphere. For an atmosphere where the temperature decreases outward monotonically the layers above are always cooler than the layers below leading to absorption of the outgoing radiation. On the other hand, where temperature increases outward (a thermal inversion) the source function contributes additional flux to the outgoing ray thereby causing emission features. Furthermore, the degree of absorption or emission is critically influenced by the magnitude of the temperature gradient as well as the opacity of the atmosphere at the particular wavelength in question. Therefore, emission spectra are strong probes of the $P$-$T$ profile of the dayside atmosphere as well as the composition. The model parametrization is similar to that of transmission spectra, with the exception that all the quantities now correspond to the 1-D dayside-averaged properties of the atmosphere. More details on thermal emission models used in retrieval can be found in various studies \citep[e.g.,][]{madhu2009,seager2010,line2013,waldmann2015b,lavie2017}. 

\subsection{Directly Imaged Spectra}

Models used in retrievals of spectra from directly imaged exoplanets are essentially the same as those of  emission spectra for transiting planets discussed above. The only difference is in the observables and, hence,  free parameters. For a directly imaged planet, only the planetary flux spectrum ($f_p$) is observed and essentially no other information about the planet is measured, including the mass and radius. As such, the planetary radius, gravity, and distance to the system are additional free parameters in the model along with the $P$-$T$ profile, chemical composition, and cloud parameters, if any. 

\section{Statistical Inference and Parameter Estimation Methods} 
\label{sec:statistical_methods}

Central to atmospheric retrieval is the parameter estimation method used to retrieve the model parameters given an observed spectrum. As discussed in previous sections, atmospheric retrieval of exoplanets is complicated by various factors including the complexity of atmospheric models, strong degeneracies between the model parameters, lack of prior knowledge, and scarcity of the data. The goal of a desired optimization algorithm is to sample a high-dimensional (10+) parameter space extensively and efficiently in search of the model solution space given the data. In order to address these various challenges, the parameter estimation methods used for atmospheric retrieval have evolved greatly over the years as discussed below and shown in table~\ref{table:retrieval_codes}.  

\subsection{From Grid-based Sampling to Bayesian Inference} 

Exoplanetary atmospheric retrieval has come a long way from grid-based sampling to detailed parameter estimation using Bayesian inference methods. The first instance of exoplanetary atmospheric retrievals \citep{madhu2009} explored a ten-dimensional (10-D) parameter space using a large grid of 10$^7$ models for each planet. Arguably, a grid of reasonable resolution in a 10-D parameter space can exceed 10$^{10}$ models, which made it computationally prohibitive even for parametric models.  Therefore, empirical metrics were used to narrow down the search volume to a more amenable 10$^7$ models. This allowed computation of contours of a goodness-of-fit statistic, e.g. a reduced $\chi^2$ , over the search volume and statistical constraints on the atmospheric parameters for a given dataset. This approach helps to obtain an empirical understanding of the model parameter space when developing a new parametric model and in conducting feasibility studies, such as limits on computational efficiency, search volume, etc. However, once a working model is established the exploration of the parameter space using such a method is both insufficient in the grid resolution and limiting in computational efficiency. Therefore, subsequent retrieval studies have investigated more formal parameter estimation methods for atmospheric retrieval while retaining the general model parametrization. 

Bayesian inference methods have gained prominence in the last decade for parameter estimation in diverse areas of astronomy, from precision cosmology \citep[e.g.,][]{tegmark2006} to exoplanet detection \citep[e.g.,][]{ford2005,eastman2013}. They allow evaluation of the full posterior distribution of the model parameters given a dataset, and prior knowledge, if any, by efficiently and comprehensively sampling the model space. Such methods are particularly useful in problems with high-dimensional and strongly degenerate parameter spaces, as is the case for exoplanetary atmospheres. Therefore, Bayesian inference methods have over the years become the mainstay of exoplanetary atmospheric retrieval codes. As shown in Table~\ref{table:retrieval_codes}, retrieval codes have incorporated a range of  Bayesian inference techniques spanning MCMC \citep{madhu2011a,benneke2012,line2013}, Optimal Estimation \citep{lee2012,line2012}, and Nested Sampling \citep{benneke2013,waldmann2015a,lavie2017,macdonald2017,gandhi2018}. 

\subsection{Bayesian Inference} 

A thorough exposition on Bayesian inference methods can be found in various sources \citep[see e.g.,][]{trotta2017} and their applications to exoplanetary atmospheric retrievals in the above studies. In what follows, we briefly discuss some key aspects. The foundation of Bayesian inference lies in the eponymous Bayes theorem which in the current context can be written as follows. 
\begin{equation}
p(\boldsymbol{\theta} | d) = \frac{p(d | \boldsymbol{\theta})~p(\boldsymbol{\theta})}{p(d)}. 
\end{equation}
Here, $\boldsymbol{\theta}$ denotes the set of parameters of an atmospheric model and $d$ denotes the data, such as an observed spectrum. $p(\boldsymbol{\theta} | d)$ is the posterior probability distribution of the model parameters given the data. $p(d | \boldsymbol{\theta})$, known as the likelihood function ($\mathcal{L}$), is the probability of the data given a parameter set. $p(\boldsymbol{\theta})$ is the prior probability distribution ($\pi$) of the model parameters independent of the data. ${p(d)}$, referred to as the evidence $\mathcal{Z}$, is the likelihood of the data marginalized over the parameter space. When considering a single model $M$, $\mathcal{Z}$ provides the normalization for the posterior. However, when considering multiple models $\{M_i\}$ with different parameterizations, the computations of $\mathcal{Z}_i$ allows model comparisons by considering the relative evidences between different models. 

The goal in Bayesian inference is to determine the posterior distributions $p(\boldsymbol{\theta} | d)$ of the model parameters for a given dataset and considering any prior knowledge of the parameters. The likelihood function ($\mathcal{L}$) determines the degree of model fit to the data for a given point in the model parameter space as  
\begin{equation}
\mathcal{L} = \mathcal{L}_0~{\rm exp} (-\chi^2/2), 
\end{equation} 
\begin{equation}
{\rm where~~} \chi^2 = \sum_i(d_i - m_i)^2/\sigma^2_i. 
\end{equation}
Here $d_i$ and $\sigma_i$ denote the mean and standard deviation of the $i$th data point, and $m_i$ is the corresponding model prediction for the given parameter set. The different Bayesian inference methods (e.g. MCMC versus Nested sampling) use different approaches to sample the model parameter space and to estimate the posterior distributions and evidences. 

\subsection{Optimal Estimation Method} 

The Optimal Estimation (OE) method has its roots in Earth-based remote sensing \citep{rodgers2000} and in retrievals of planetary atmospheres in the solar system \citep{irwin2008}. More recently, it has been applied to retrievals of exoplanetary atmospheres \citep{lee2012}. The method involves optimizing the likelihood function using a non-linear least squares minimization scheme such as the Levenberg-€"Marquardt algorithm. The OE method allows specification of priors for the parameters in the cost function, assuming a Gaussian-distributed prior covariance matrix. This is particularly relevant for Earth based retrievals where prior values of parameters can be approximated based on direct measurements \citep{irwin2008}. The advantage of this method is that only a small number of iterations are required to obtain a model fit to the spectral data and is known to converge efficiently when high resolution and high signal-to-noise (SNR) data are available, e.g., in spectral retrievals of Earth and solar system objects. 

The OE method is somewhat limited for large and multi-modal parameter spaces with strong degeneracies and non-Gaussian posterior distributions, which is typically the case for exoplanetary atmospheres. The method has been shown to be inaccurate for low-resolution low-SNR data as common for current exoplanetary spectra \citep{line2013}, but in the limit of high-resolution high-SNR data it approaches the performance of more sophisticated Bayesian methods discussed below for single-modal parameter spaces. The OE method assumes Gaussian distributed uncertainties in the model parameters and uses gradient-descent optimization which is arguably less efficient in detecting global minima and sampling multi-modal spaces compared to Monte Carlo methods such as MCMC or Nested sampling. Nevertheless, the OE method has been used in the retrievals of several exoplanetary spectra to provide important constraints on their atmospheric properties. \citep{lee2012,lee2014,barstow2017}. 

\subsection{Markov chain Monte Carlo (MCMC)} 

The MCMC method is one of the most widely used Bayesian inference methods in astronomy \citep{trotta2017}. In this method the exploration of the parameter space starts at an initial guess and progresses as a random walk wherein any given step in the ``chain" depends only on the previous step. For example, in the commonly used Metropolis-Hastings algorithm the progression of the random walk is guided by the following procedure. At each step in the chain the decision to accept the next step is based on the ratio of the posteriors between the two steps. Consider a current step with parameter set $\theta_i$ giving the posterior $p(\boldsymbol{\theta_i} | d)$. The parameter set for next step $\theta_{i+1}$ is drawn from a pre-specified distribution, such as a Gaussian centered on the current step with a pre-specified variance (``jump-length") for each parameter, and has a posterior $p(\boldsymbol{\theta_{i+1}} | d)$. Then, step i+1 is accepted with a probability $ p = {\rm min}(p(\boldsymbol{\theta_{i+1}} | d)/p(\boldsymbol{\theta_{i}} | d), 1)$ such that step i+1 is accepted if $p$ is greater than a random number drawn from a uniform distribution between 0 and 1. The resulting full chain of steps through the parameter space essentially gives the joint posterior probability distribution of all the model parameters. More details on the MCMC method and variations thereof can be found in various works \citep{tegmark2004,ford2006,line2013,trotta2017}. 

The MCMC method has been extensively used for exoplanetary atmospheric retrievals \citep{madhu2011a,benneke2012,line2013}. The method allows efficient and extensive sampling of the posterior distribution, and allows the specification of prior distributions of the parameters where applicable. Generally for exoplanetary atmospheres there is usually very little prior knowledge. Therefore, retrievals typically allow for uniform priors with conservative ranges. Despite its capabilities, the MCMC method faces some limitations, especially for complex parameter spaces. For example, the MCMC method is not optimized for calculating the evidence $\mathcal{Z}$ which is computationally demanding. This is acceptable for parameter estimation of a given model, where $\mathcal{Z}$ acts as a normalization constant, but makes it challenging to conduct model comparisons when multiple models are plausible. Secondly, the MCMC method requires the user to specify the width of the distribution from which to draw each subsequent step, which is judged by trial and error and can effect convergence. These limitations are alleviated in the Nested Sampling method discussed below. 

\subsection{Nested Sampling} 

Recently, the Nested Sampling (NS) method has emerged as a powerful alternative to the MCMC method in Bayesian inference \citep{skilling2006,shaw2007,feroz2009}. As such, it has been promptly adopted in exoplanetary atmospheric retrieval codes \citep[e.g.,][]{benneke2013,line2015,waldmann2015a,lavie2017,macdonald2017,gandhi2018}. The NS method is also a Monte Carlo method like the MCMC but with a different approach to sample the parameter space. Instead of starting with an initial guess and following a Markov chain, as in MCMC, the NS method starts with a given number of ``live points" in the parameter space randomly drawn from the prior distribution \citep{feroz2009}. At each step the point with the lowest likelihood ($\mathcal{L}_{\rm min}$) is discarded and replaced by another point drawn from the prior distribution with $\mathcal{L}>\mathcal{L}_{\rm min}$, i.e. from the prior volume contained within the iso-likelihood contour of $\mathcal{L}_{\rm min}$. Thus, the live points are drawn from progressively shrinking ellipsoids bound by the iso-likelihood contours in subsequent trials. The process is repeated as the contours sweep through the parameter space and the evidence $\mathcal{Z}$ is calculated until a pre-set tolerance on the fractional change in $\mathcal{Z}$ is reached. The required tolerance ensures that convergence is naturally reached. Once the evidence is determined, the posterior distributions are computed based on all the points sampled in the parameter space over the entire optimization process. 

The NS method has several advantages over other Bayesian inference methods. One of the main advantages of the NS method is that it is designed to be highly efficient for computing the Bayesian evidence ($\mathcal{Z}$) for a given model, which makes it particularly desirable when comparing between multiple models. In efficiently exploring the model parameter space to compute $\mathcal{Z}$ with high accuracy, the NS method also naturally allows high density sampling of the posterior distribution. This makes the NS method especially suited for handling complex model parameter spaces with multimodal and non-Gaussian posterior distributions. Secondly, the optimisation algorithm is naturally parallelised, thereby significantly reducing computation time. Finally, unlike MCMC, it does not require specification of the distribution from which to sample the parameters which would be required in each step of a Markov Chain. 

\begin{figure*}[hbt!]
\centering
\includegraphics[width=\textwidth]{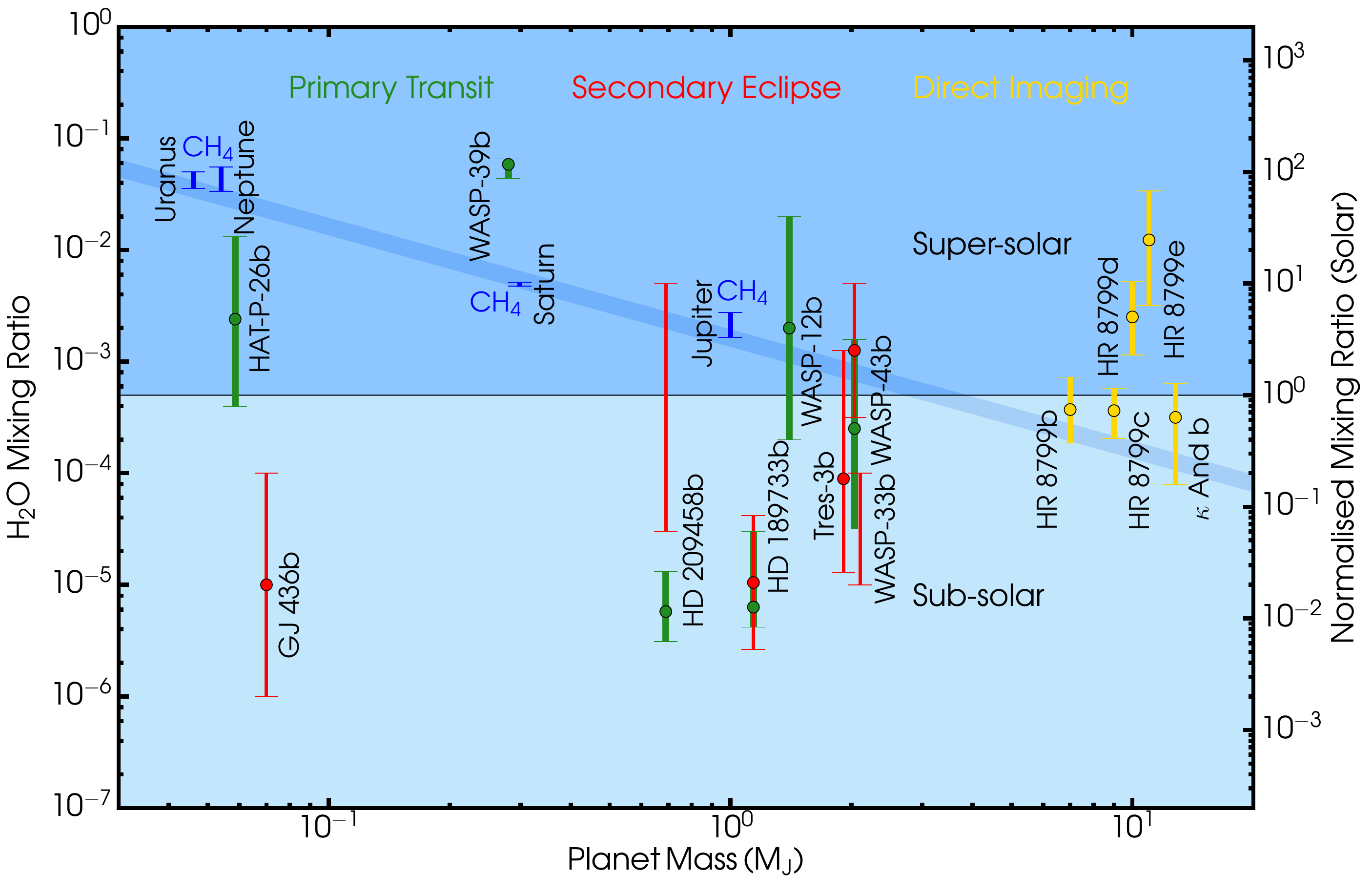}
\caption{Retrieved H$_2$O abundances for exoplanets in the literature with uncertainties in H$_2$O mixing ratios below 2 dex. The abundances derived from spectra obtained using different methods are colour-coded as shown at the top. Also shown (in blue) are the CH$_4$ abundances for the four solar system giant planets for which the H$_2$O abundances are not known. The solar system CH$_4$ abundances are obtained for Jupiter and Saturn from \citet{atreya2016},\citet{wong2004}, and \citet{fletcher2009}, for Neptune from \citet{karkoschka2011}, and for Uranus from \citet{sromovsky2011}. The exoplanet H$_2$O abundances are from the various works: HD~209458b \citep{madhu2014b,macdonald2017,line2016b}, HD~189733b \citep{madhu2014b,waldmann2015b}, WASP-12b \citep{kreidberg2015}, WASP-43b  \citep{kreidberg2014a}, WASP-33b \citep{haynes2015}, TrES-3 \citep{line2014}, GJ~436b \citep{moses2013a}, HAT-P-26b \citep{wakeford2017}, WASP-39b \citep{wakeford2018}, HR~8799 planets \citep{lavie2017}, $\kappa$ And b \citep{todorov2016}.}
\label{fig:retrievals}
\end{figure*}

\section{Results} 

The retrieval methods described above have been used to retrieve chemical abundances, temperature profiles, and other atmospheric parameters for a number of exoplanets using different observational methods. Here we discuss the constraints reported in the literature for transiting planets observed via transmission and emission spectra as well as directly imaged planets in emission spectra.  

\subsection{Transmission spectra of transiting exoplanets} 
Most of the retrievals to date have been conducted on transmission spectra. A transmission spectrum probes the atmosphere at the day-night terminator region of a transiting exoplanet. The retrieved properties include chemical compositions, pressure-temperature (P-T) profiles, and cloud properties. In general, a transmission spectrum is less sensitive to detailed shape of the P-T profile but still provides a reasonable constraint on the representative photospheric temperature at the day-night terminator. On the other hand, the transmission spectrum is highly sensitive to the composition and presence of clouds/hazes. In what follows, we discuss published statistical constraints on atmospheric properties of exoplanets using retrievals of transmission spectra.  

\subsubsection{Key sources of data} Here we focus on reported high-precision transmission spectra with instruments whose systematics are well characterized and results reproducible. Datasets used in initial retrieval studies were limited by large uncertainties and/or underestimated systematics rendering the abundance determinations unreliable. The advent of the HST WFC3 spectrograph \citep{mccullough2012} has truly opened the era of high-precision abundance measurements from transmission spectra of exoplanets. The HST WFC3 G141 grism with its spectral range of 1.1-1.7 $\mu$m contains strong absorption features due to H$_2$O, along with other molecules (e.g. CH$_4$, NH$_3$, HCN). Additionally, the WFC3 instrument has proven to be highly stable and conducive to transit spectroscopy with demonstrated understanding of the systematics in numerous studies \citep[e.g.,][]{deming2013,mandell2013,mccullough2014,kreidberg2014b}. Besides HST WFC3 in the near-infrared, the HST STIS spectrograph in the visible is sensitive to optical slopes of transmission spectra which in turn constrain sources of scattering (e.g. aerosols or molecular Rayleigh scattering) in the atmospheres \citep[e.g.,][]{pont2008,pont2013,sing2016,wakeford2015,pinhas2017}. Other data sources used in retrievals include photometric data in the infrared obtained with the Spitzer Space Telescope typically at 3.6 $\mu$m and 4.5 $\mu$m \citep[e.g.,][]{fraine2014} as well as ground-based spectra/photometry in the visible and near-infrared \citep[e.g.,][]{sedaghati2017}. 

\subsubsection{Reporting retrieved abundances} The retrieved chemical abundances are typically reported as volume mixing ratios (i.e. number density of a species relative to total or relative to H$_2$ which is the dominant species in giant planet atmospheres). It is also common to refer to the mixing ratios relative to ``solar" values, i.e., those expected in thermochemical equilibrium at the relevant temperature ($T$) for an atmosphere with solar elemental abundances, O/H = 5 $\times$ 10$^{-4}$, C/H = 2.5 $\times$ 10$^{-4}$, C/O = 0.5 \citep{asplund2009}. The portion of oxygen in H$_2$O in chemical equilibrium depends primarily on the overall metallicity, the C/O ratio, and the temperature and pressure \citep[see e.g.,][for a detailed review on atmospheric chemistry]{madhu2016}. For solar abundance atmospheres with $T$ $\sim$ 1200-3500 K and $\lesssim$1200 K, the typical value of solar H$_2$O/H$_2$ is 5$\times$ 10$^{-4}$ and $\sim$10$^{-3}$, respectively, with the remaining oxygen locked in CO and other species. Reported H$_2$O abundances greater or lower than these values are referred to as super-solar or sub-solar H$_2$O, respectively. 

\subsubsection{Abundance estimates in hot Jupiters} The majority of retrieved abundances using HST WFC3 transmission spectra have been reported for H$_2$O in hot Jupiters, as shown in Fig.~\ref{fig:retrievals}. The earliest high-precision WFC3 transmission spectra were observed for the hot Jupiters HD~209458b \citep{deming2013} and HD~189733b \citep{mccullough2014}. Atmospheric retrievals of these spectra \citep{madhu2014b} suggested sub-solar H$_2$O abundances in these hot Jupiters, assuming cloud-free atmospheres. Other notable examples of high-precision transmission retrievals include the hot Jupiters WASP-43b \citep{kreidberg2014b} and WASP-12b \citep{kreidberg2015} for which H$_2$O abundances between 0.1-3$\times$ solar have been derived. H$_2$O abundance estimates in giant exoplanets today are routinely achieving uncertainties below 1 dex, as shown in Fig.~\ref{fig:retrievals}, which is a significant achievement considering that the true H$_2$O abundance is not known for any of the giant planets in the solar system owing to their low temperatures \citep{atreya2016,madhu2016}. 

The precision of an abundance estimate is directly related to the quality of the observed spectrum. In particular the two factors of an observed spectrum that affect critically are (a) the precision, and (b) the spectral range. In particular the availability of data spanning the visible (e.g. using HST STIS) to infrared (e.g. HST WFC3 and Spitzer) range is critical for accurate retrievals of transmission spectra. The best example is the visible-IR transmission spectrum of HD~209458b where the average precision on the data is $\sim$25 ppm in the WFC3 which led to an abundance estimate with an uncertainty of $\lesssim$ 0.5 dex \citep{madhu2014b,barstow2017,macdonald2017}. Such a dataset was feasible for HD~209458b considering that it orbits one of the brightest exoplanet host stars (V = 7.8 magnitude). Fainter host stars often require integration of multiple spectra to achieve similar precisions \citep{stevenson2014c,kreidberg2014b}. Thus, focused repeat observations with HST have the potential to yield high-precision abundance estimates in more giant exoplanets. The upcoming JWST will substantially revolutionise abundance determinations in exoplanet atmospheres with a much larger aperture and spectral range \citep{greene2016}.

Beyond H$_2$O abundances, constraints on other atmospheric properties from retrievals of transmission spectra are relatively sparse. In recent years, thanks to the combination of near-infrared and optical spectra, there have been nominal constraints on the parameters of clouds/hazes in the atmospheres, such as cloud top pressure, optical slopes due to hazes, patchiness, Na/K, etc \citep{barstow2017,macdonald2017}. Recent retrievals are also suggesting the first indications of Nitrogen-based chemistry in the form of NH$_3$ and/or HCN \citep{macdonald2017b}. Other recent developments in transmission retrievals include the detection of TiO in the transmission spectrum of hot Jupiter WASP-19b \citep{sedaghati2017}.

A recurring finding in the majority of transmission spectra of hot Jupiters is that the amplitudes of H$_2$O features in the spectra are significantly muted \citep[e.g.,][]{deming2013,mccullough2014,mandell2013,kreidberg2015}. The features contain the equivalent of $\sim$2 atmospheric scale heights compared to 5-10 expected for a saturated molecular absorption feature \citep[e.g.][]{madhu2015}. It has been argued that such muted features can be caused by either clouds/hazes in the atmospheres obstructing parts of the atmospheres from view \citep{deming2013,sing2016} or due to inherently low H$_2$O abundances in the atmospheres \citep{madhu2014b}. In a recent study \citet{sing2016} reported broadband transmission spectra of ten hot Jupiters and, using a forward model grid, suggested that a diversity of clouds/hazes with solar or super-solar H$_2$O abundances could explain all the spectra. However, a subsequent retrieval study \citep{barstow2017} reported the contrary that almost all the hot Jupiters in the \citet{sing2016} sample indicated sub-solar H$_2$O abundances, consistent with other retrieval studies \citep{madhu2014b,macdonald2017} for some of the planets. This demonstrates the importance of retrievals over traditional equilibrium models in deriving abundances. On the other hand, transmission retrievals of some hot Jupiters, e.g. WASP-43b and WASP-12b \citep{kreidberg2014a,kreidberg2015}, while also consistent with sub-solar H$_2$O abundances have uncertainties large enough to allow somewhat super-solar abundances. 

\subsubsection{Abundance estimates in hot Neptunes and super-Earths}
In recent years, we are beginning to witness detections of molecular features in transmission spectra of extrasolar ice giants, the so called hot Neptunes. Early HST WFC3 observations of hot Neptunes \citep[e.g., GJ~436b,][]{knutson2014a} and super-Earths \citep[e.g., GJ~1214b,][] {kreidberg2014a} showed mostly flat transmission spectra. However, more recent observations of hot Neptunes HAT-P-11b \citep{fraine2014} and HAT-P-26b \citep{wakeford2017} have shown clear absorption features of H$_2$O at high significance. Retrievals of these spectra have reported abundances in terms of atmospheric metallicity with 1-$\sigma$ constraints of $\sim$40-300$\times$ solar for HAT-P-11b and $\sim$1-26$\times$solar for HAT-P-26b. Note that here H$_2$O is assumed to be the primary carrier of oxygen whose abundance in turn is used to represent the metallicity. Besides these abundance constraints currently there is still a dearth of precise abundance constraints for hot Neptunes and super-Earths. The imminent arrival of JWST is expected to revolutionize abundance estimates in such low mass planets. 

\subsection{Thermal emission spectra of transiting exoplanets} 

Emission spectra of transiting exoplanets provide constraints on the properties of their dayside atmospheres. Unlike transmission spectra, emission spectra are strongly sensitive not only to the chemical composition but also to the $P$-$T$ profile of the atmosphere. It is thermal emission directly from the planet that is measured and is strongly governed by the temperature distribution in the atmosphere. Emission measurements in one or more bands have been measured for over 50 exoplanets to date. However, in order to conduct detailed retrieval of the atmospheric properties, observations over a long spectral range are required. Such observations are available for $\lesssim$10 planets to date. 

Thermal emission retrievals have provided key constraints on three properties of the dayside atmospheres of hot transiting exoplanets: molecular abundances, C/O ratios, and thermal inversions. The constraints have been reported mostly for hot Jupiters but also include one hot Neptune. The contributions of these properties in different observational bandpasses are discussed in several works \citep[e.g.][]{burrows2008,madhu2012,madhu2014a,moses2013b,molliere2016}. 

\subsubsection{Molecular Abundances}

The chemical abundances have been reported mainly for H$_2$O with broad constraints on a few other species such as CO, CO$_2$, and CH$_4$. Figure~\ref{fig:retrievals} shows the H$_2$O abundance constraints from various studies. The earliest abundance estimates in emission retrievals were based primarily on broadband Spitzer photometry \citep{madhu2009,stevenson2010,madhu2011b,lee2012,line2013} and constrained molecular abundances with large uncertainties. Additionally, the most extensive Spitzer observations used in such retrievals, e.g. for hot Jupiters HD~189733b \citep{charbonneau2008} and HD~209458b \citep{knutson2008} have since been revised to very different values \citep[e.g.][]{knutson2012,diamond-lowe2014} rendering the former abundances unreliable. The success of HST WFC3 and ground-based near-infrared photometry in recent years have led to more reliable constraints on H$_2$O abundances. Such reported H$_2$O abundances on the daysides of hot Jupiters include 0.4-3.5$\times$ solar in WASP-43b \citep{kreidberg2014b}, $\lesssim$0.01$\times$ solar in WASP-12b \citep{madhu2011a,stevenson2014a}, 0.06-10$\times$ solar in HD~209458b \citep{line2016b}, $\lesssim$ 0.2 $\times$ solar in WASP-33b \citep{haynes2015}, among others. 

The sum-total of abundance measurements convey two key findings. Firstly, with the combination of high-precision emission data from HST, Spitzer, and ground-based instruments we are now able to measure H$_2$O abundances with uncertainties to within an order of magnitude. Secondly, similar to transmission spectra, the abundance estimates from emission retrievals are also suggesting H$_2$O abundances lower than originally expected. The median values of the H$_2$O abundances for most of the planets reported to date are sub-solar, though the uncertainties allow between 0.1-3.5$\times$ solar values. Additionally, the incidence of clouds/hazes which are degenerate with the H$_2$O abundance determinations in transmission spectra is less of a problem in emission retrievals, given the much higher dayside temperatures and the fact that the spectrum probes the temperature profile and composition above a putative cloud deck if any. 

Abundance retrievals for species other than H$_2$O are significantly less precise. Nevertheless, several retrieval studies have reported useful limits on carbon-bearing species with potentially interesting implications. For example, significantly super-solar abundances of CO and CO$_2$ have been reported in the hot Neptune GJ~436b suggesting strong non-equilibrium chemistry and high ($\gtrsim$ 10-30$\times$solar) overall metallicity \citep{stevenson2010,madhu2011b,lanotte2014}. 

\subsubsection{C/O Ratios }

An important development in emission retrievals is the feasibility of constraining C/O ratios in exoplanetary atmospheres. This was first demonstrated for the hot Jupiter WASP-12b using MCMC retrievals on a multi-band photometric dataset spanning a wide spectral range from 1-10 $\mu$m comprising Spitzer and ground-based photometry \citep{madhu2011a}. The retrieved abundances suggested low H$_2$O ($\lesssim$0.01$\times$ solar) and higher CO and CH$_4$ compared to H$_2$O leading to a C/O $\ge$ 1, which is significantly carbon-rich compared to the solar C/O of 0.54 with important implications \citep{madhu2011c,madhu2012}. The claim of high C/O in WASP-12b has been contested in subsequent studies \citep[e.g.,][]{crossfield2012,cowan2012} though a recent analysis considering newer Spitzer and HST observations  reinstated the claim \citep{stevenson2014a}, especially due to the non-detection of H$_2$O in the WFC3 emission spectrum. On the other hand, a transmission spectrum of the terminator shows a low-amplitude H$_2$O feature that is consistent with a oxygen-rich as well as a carbon-rich composition \citep{kreidberg2015}. The dependance of H$_2$O abundance on the C/O ratio varies with temperature; for C/O = 1 the H$_2$O abundance is more strongly depleted at higher temperatures typical of dayside atmospheres \citep{madhu2012}. Nevertheless, the C/O ratio of WASP-12b is currently still under debate, considering the limited data available, which can be resolved by future observations with the upcoming JWST. 

The case of WASP-12b demonstrated the potential of thermal emission spectra to constrain C/O ratios in exoplanetary atmospheres. It is to be noted, however, that reliable constraints on C/O ratios are possible only when observations over a long spectral baseline are available with the observed bands containing spectral features from both H$_2$O as well as prominent carbon-bearing species such as CO and/or CH$_4$. This is ostensibly possible with the combination of spectra with HST WFC3 and the Spitzer IRAC bands between 3-8 $\mu$m. However, few systems have such extensive datasets. Using the limited datasets available, C/O ratios for several hot Jupiters have been reported to lie between 0.1-1 \citep[][]{madhu2012,line2014,haynes2015,kreidberg2014b,line2016b}. On the other hand, when adequate data are not available and only H$_2$O abundances can be measured from HST WFC3, some studies have also attempted to infer the C/O ratio based on assumptions of thermochemical equilibrium \citep{kreidberg2015,line2016b}. Overall, the C/O ratio has emerged to be a  measurable quantity of high importance for future higher resolution spectroscopy, e.g., with JWST. 

\subsubsection{Thermal Inversions}

Thermal emission spectra have also been instrumental in constraining temperature profiles in the dayside atmospheres of exoplanets. In particular, the search for thermal inversions in hot Jupiter atmospheres has been one of the most important pursuits in exoplanetary atmospheres in the past decade. The temperature gradient in the atmosphere along with the composition, through opacity, governs the amplitude of the spectral features. Whereas a temperature profile monotonically decreasing outward gives rise to absorption features, that increasing outward (i.e. a thermal inversion) gives rise to emission features. On the other hand, an isothermal atmosphere emits as a blackbody with the corresponding temperature. These features have been predicted using self-consistent models well before retrievals came into practice \citep[e.g.,][]{hubeny2003,burrows2007,fortney2008}. 
 
Retrievals of emission spectra of transiting hot Jupiters have revealed the diversity in their $P$-$T$ profiles. Initial retrievals \citep{madhu2009,madhu2010} revealed the strong degeneracies between temperature profiles and compositions. Thanks to recent HST WFC3 spectra and ground-based photometry of thermal emission, strong constraints on temperature profiles have become possible for several transiting exoplanets. Most of these planets observed to date have been found to host no thermal inversions \citep{madhu2010,line2013}, some highlights being WASP-43b \citep{kreidberg2014b,stevenson2014c}, HD~209458b \citep{line2016b}, and WASP-12b \citep{madhu2011a,stevenson2014a}. On the other hand, thermal inversions have been convincingly detected in three hot Jupiters: WASP-33b \citep{haynes2015}, WASP-121b \citep{evans2017} and WASP-18b \citep{sheppard2017}, which are amongst the most extremely irradiated hot Jupiters known, with equilibrium temperatures of $\sim$3000 K. Therefore, retrievals of emission spectra to date show that thermal inversions are prevalent in only the hottest of hot Jupiters at temperatures of $\sim$3000 K, much above the $\sim$1500 K boundary originally suggested by \cite{fortney2008}. 

\subsection{Thermal emission spectra of directly imaged planets} 

Retrievals of directly-imaged planets are still in their infancy considering the small number of such planets with atmospheric spectra. On the one hand, the spectra are typically of much higher quality than those currently available for transiting planets \citep{barman2011,konopacky2013}. On the other hand, direct imaging retrievals are more challenging compared to transiting planet retrievals because of the various unknowns and degeneracies. For example, for directly imaged planets only the emission spectrum of the planet is observed without much a priori  information about several of the system parameters, e.g. radius, gravity/mass, any measure of temperature, distance to the system, etc. Therefore, all these quantities need to be set as free parameters in the retrievals. And, considering that these are typically young systems, the luminosity and radius of the planet are strongly dependent on the age and can have a wide range\citep{burrows2001}. Finally, given the low levels of irradiation, such planets have been known to host significantly dusty atmospheres \citep{marley2012} and convection playing a stronger role in the temperature profiles compared to transiting planets which are highly irradiated. 

Some recent studies have made important advancements towards retrievals of directly imaged planets with important constraints on their atmospheric properties. Retrievals have been reported for the planets in the HR 8799 system \citep{lee2013,lavie2017} and for the $\kappa$ {Andromedae} b \citep{todorov2016}, using different retrievals approaches: NEMESIS Optimal  Estimation \citep{lee2013}, MULTINEST  Nested sampling \citep{lavie2017}, and MCMC \citep{todorov2016}. \citet{lee2013} and \citet{lavie2017} reported constraints on the abundances of the prominent molecules H$_2$O, CO, CH$_4$, and CO$_2$ in HR 8799b, with the later study also reporting constraints on a subset of the species in HR 8799c,d,e. Generally, the studies find super-solar abundances in all the species and super-solar C/O ratios ($\sim$0.75-0.96) for HR~8799b and c. And, \citet{lavie2017} report sub-solar C/H and C/O and super-solar O/H for HR~8799 d and e. On the other hand, \citet{todorov2016} report the H$_2$O abundance of $\kappa$ {Andromedae} b to be nearly solar, albeit with a larger uncertainty. These studies demonstrate the high-precisions with which abundances can be retrieved for directly-imaged planets. On the other hand, the studies also reveal the challenges in such retrievals. For example,  \citet{lee2013} find too small radii for young giant planets (0.6-0.8) that are seemingly unphysical whereas \citet{lavie2017} had to adopt a strong prior on the radii (1.2 $\pm$ 0.1 R$_J$), log(gravity) (4.1 $\pm$ 0.3), and distance (39.4 $\pm$ 1 pc) to facilitate the retrievals.  

\section{Challenges and Future Directions} 

Retrieval is an extremely powerful tool but also one to be used with due  care. Inherent to Bayesian inference is the grand reality that in the limit of very poor data quality the posterior asymptotes to the prior. Therefore, arguably the best approach in retrieval is to allow the data to shepherd one to the reality of an atmosphere with as few model assumptions as absolutely necessary. On the other hand, the insufficient data quality means that one is tempted to recourse to arguments of physical/chemical plausibility to narrow down the solution space, leaving open the definition of plausibility in the unknown conditions of an exoplanetary atmosphere. Furthermore, given the highly complex, non-linear, and degenerate parameter space of atmospheric models, any statistical inference algorithm used must ensure to sample the parameter space rigorously and efficiently. The main limitation at the moment is the data quality, of a limited spectral range and precision, which is expected to greatly improve with the upcoming JWST and large ground-based facilities. In what follows, we discuss some key challenges faced in retrievals of exoplanetary spectra with current data in the hope that they serve as important lessons for the future. 

\subsection{Degeneracies in Abundance Estimates} The main challenge in abundance estimates using retrievals is the prevalence of strong degeneracies between different chemical species and among other atmospheric properties. Foremost among them is the degeneracy between clouds/hazes and abundances, particularly while retrieving transmission spectra. A low amplitude spectral feature can be caused either by clouds/hazes \citep{deming2013,sing2016} or due to inherently low abundances \citep{madhu2014b}. Secondly, in transmission retrievals the abundances are also degenerate with the reference pressure corresponding to the quoted radius of the planet if spectra in only a limited infrared spectral range, such as HST WFC3, are available \citep{heng2017}. Both these degeneracies can be resolved with high-precision spectra over a long spectral baseline, from visible to infrared, and with multiple molecular features, as demonstrated in some recent studies \citep{barstow2017,macdonald2017}. While such data are available only for a handful of planets currently, these examples highlight the critical importance of optical spectra along with infrared spectra in constraining molecular abundances. 

\subsection{Low-amplitude Spectral Features} One of the greatest surprises in transit spectroscopy in recent years is the ubiquitously low amplitudes in the spectra. As discussed in the results section, even for the hottest of hot Jupiters the spectra are surprisingly muted, amounting to $\lesssim$2 scale heights instead of 5-8 expected for a saturated absorption feature. This is seen in both transmission as well as emission spectra. The low amplitudes in transmission spectra could be due to clouds or low H$_2$O abundances. Retrievals of current broadband data with models including clouds/hazes still suggest sub-solar abundances as the favored explanation for hot Jupiters \citep{barstow2017,macdonald2017} but it remains to be further investigated with future data. On the other hand, the low amplitudes in emission spectra could potentially be due to low temperature gradients or, again, low abundances. Whatever the present interpretation, the ground reality is that spectral observations and retrievals using future facilities such as JWST should be prepared for low amplitude spectral features in potentially most planets. 

\subsection{Biases in Estimating C/O ratios and Metallicities } When data are insufficient to constrain a certain quantity, retrievals sometimes invoke additional model assumptions to narrow down the solution space. While this could be useful to some extent to rule out extremely unphysical solutions, caution must be exercised on this path. For example, this is commonly seen in quoting constraints on quantities such as the C/O ratio and metallicities \citep{kreidberg2015,line2016b,wakeford2017} when only HST WFC3 data are available with constraints on only the H$_2$O abundance. Retrievers then assume chemical equilibrium, amongst other factors, to estimate which C/O ratios could cause the retrieved  H$_2$O abundances for the retrieved $P$-$T$ profiles; the process is sometimes termed ``chemically consistent retrievals". Similarly, H$_2$O is often used as a proxy for metallicity by inherently assuming a certain C/O ratio. At the extreme end of this trend, inferences of metallicities are sometimes made solely using equilibrium forward models, i.e., without retrievals \citep{sing2016}, only to find contrasting conclusions when retrievals are performed later on the same datasets \citep{barstow2017}. Whereas \citet{sing2016} claimed none of the ten planets in their survey were consistent with sub-solar H$_2$O, \citet{barstow2017} reported that all the planets were consistent with sub-solar H$_2$O. Therefore, care must be taken against using such inferences of C/O ratios or metallicities with equilibrium assumptions as a true measure of the atmospheric composition since various non-equilibrium effects are unaccounted for in the process. Ultimately, the most reliable estimates are those that are obtained from retrievals with minimal model assumptions and guided by high-fidelity datasets. 

\subsection{New Trends and Future Prospects} 

New advancements in retrievals are being made in several directions. With increasing data quality there is increasing sophistication in the forward models used in retrievals. One development in this direction is the consideration of two-dimensional effects, such as multiple P-T profiles \citep{feng2016} and inhomogeneous cloud cover\citep{line2016a,macdonald2017}. Secondly, we are also seeing the emergence of ``hybrid" codes where retrievals are fully interfaced with self-consistent equilibrium models to place constraints on disequilibrium phenomena, such as departures from radiative-convective and thermochemical equilibria \citep{gandhi2018}. Thirdly, on the observational side, a major development is the idea of combining low-resolution transit spectroscopy with very high resolution (R$\sim$10$^5$) ground-based spectroscopy\citep{brogi2017}. The latter method, which is essentially doppler spectroscopy of molecular lines in the planetary atmosphere, has in recent years proved to be the most effective technique for conclusive detections of molecular species in exoplanetary atmospheres \citep{snellen2010,brogi2012,birkby2017}. Combining this method with traditional transit spectroscopy provides a new avenue for high-fidelity atmospheric retrievals. There have also been efforts to try machine learning techniques such as artificial neural networks for retrievals \citep{waldmann2016} but their efficacy on real datasets and benefits over state-of-the-art Bayesian inference methods remains to be seen. 

We are at the beginning of a revolution in atmospheric characterization of exoplanets. Atmospheric retrievals today are equipped with state-of-the-art Bayesian inference techniques combined with detailed forward models and are limited only by current data quality.  The latter is going to change very soon with the imminent arrival of JWST and ramping up of ground-based spectroscopy with large facilities. JWST will have a broad spectral range ($\sim$0.6-25 $\mu$m), high spectral resolution and precision. These capabilities will allow detection of a wide range of species besides H$_2$O, such as CO, CO$_2$, CH$_4$, NH$_3$, and others, and precise determinations of their abundances \citep{greene2016}, along with constraints on the $P$-$T$ profiles, aerosols forming clouds/hazes, energy budget, etc. These constraints in turn will allow us to understand a wide range of atmospheric processes such as non-equilibrium chemistry, thermal inversions, atmospheric dynamics, cloud formation, etc. The molecular abundances will also allow precise constraints on the C/H, O/H, C/O, and other elemental abundance ratios, which will be instrumental in constraining planetary formation conditions \citep[see e.g. review by][]{madhu2016}. And, finally, atmospheric retrievals of low mass planets could pave the way to the first detections of biosignatures epitomizing the holy grail of the exoplanetary science. 

\begin{acknowledgement}
The author acknowledges the tireless efforts by various groups working on exoplanetary atmospheric retrieval which has led to the exponential rise in this area in the last eight years. The author thanks A. Pinhas for help with Table 1 and Fig. 3, A. Pinhas and R. MacDonald for help with references, and L. Welbanks for help with Fig. 2. The author thanks the chapter editor Sara Seager for very helpful comments. 

\end{acknowledgement}

\bibliographystyle{spbasicHBexo}  
\bibliography{refs_retrieval.bib} 

\end{document}